\documentclass[11pt,preprint]{aastex}
\usepackage{graphics,graphicx,epsfig,lscape}

\usepackage{amssymb}
\usepackage{subfigure}
\usepackage{epstopdf}
\usepackage{natbibspacing}
\usepackage{natbib}
\newcommand{\msun}{M$_{\odot}$ }

\newcommand{\ie}{\textit{i.e.}~}
\newcommand{\GALEX}{\textit{GALEX}}

\begin{document}

\title{An OSIRIS study of the gas kinematics in a sample of UV-selected galaxies: Evidence of ``Hot and Bothered'' starbursts in the local Universe}
\author{
Antara R. Basu-Zych\altaffilmark{1},
Thiago S. Gon\c calves\altaffilmark{2},
Roderik Overzier\altaffilmark{3},
David R. Law\altaffilmark{5},
David Schiminovich\altaffilmark{1},
Tim Heckman\altaffilmark{4},
Chris Martin\altaffilmark{2},
Ted Wyder\altaffilmark{2}, 
Matt O'Dowd\altaffilmark{1}}
\altaffiltext{1}{Department of Astronomy, Columbia University, 550 West 120th Street, New York, NY 10027; antara,ds, matto@astro.columbia.edu}
\altaffiltext{2}{California Institute of Technology, MC 405-47, 1200 East California Boulevard, Pasadena, CA 91125; tsg@astro.caltech.edu, cmartin, wyder@srl.caltech.edu}
\altaffiltext{3}{Max-Planck-Institut f\"{u}r Astrophysik, D-85748 Garching, Germany; overzier@MPA-Garching.MPG.DE}
\altaffiltext{4}{Center for Astrophysical Sciences, The Johns Hopkins` University, 3400 N. Charles St., Baltimore, MD 21218; heckman@pha.jhu.edu}
\altaffiltext{5}{Hubble Fellow.  Department of Physics and Astronomy, University of California, Los Angeles, CA 90095; drlaw@astro.ucla.edu}

\begin{abstract}
We present data from Integral Field Spectroscopy for 3 supercompact UV-Luminous Galaxies (ScUVLGs). As nearby ($z\sim 0.2$), compact (R$_{50}\sim$1-2 kpc), bright Paschen-$\alpha$ sources, with unusually high star 
formation rates (SFR$=$3$-$100 \msun yr$^{-1}$), ScUVLGs are an ideal population for studying detailed kinematics and dynamics in actively star-forming galaxies. In addition, ScUVLGs appear to be excellent analogs to high redshift Lyman Break Galaxies (LBGs) and our results may offer additional insight into the dynamics of LBGs. Previous work by our team has shown that the morphologies of these galaxies exhibit tidal features and companions,   and in this study we find that the dynamics of ScUVLGs are dominated by disturbed kinematics of the emission line gas-- suggestive that these galaxies have undergone recent feedback, interactions or mergers. While 2 of the 3 galaxies do display rotation, v/$\sigma<$1 -- suggesting dispersion dominated kinematics rather than smooth rotation.  We also simulate how these observations would appear at $z\sim$2. Lower resolution and loss of low surface brightness features causes some apparent discrepancies between the low-$z$ (observed) and high-$z$ (simulated) interpretations  and quantitatively gives different values for v/$\sigma$, yet simulations of these low-$z$ analogs manage to detect the brightest regions well and resemble actual high-$z$ observations of LBGs. 
\end{abstract}

\section{Introduction}\label{sec:intro}
Across cosmic time, processes driving galaxy formation are expected to transition from hierarchical merging at early times to secular evolution at low redshift \citep{Whiterees, abadi, kormendy}. In the case of hierarchical merging, galaxies grow as halos merge, causing disturbed kinematics and dynamics. Secular evolution describes slow and steady mechanisms, including gas accretion and internal processes, and is consistent with smooth rotation. 

The Galaxy Evolution Explorer (\GALEX) has uncovered a population of local ($z=0.1-0.3$), intensely star-forming galaxies. These ultraviolet luminous galaxies (UVLGs) are selected based on their Far-UV (FUV) luminosities (L$_{FUV}$ $>$ 2$\times$10$^{10} \rm L_{\odot}$). UVLGs exhibiting the highest surface-brightnesses (I$_{FUV} >10^{9}$ L$_{\odot}$ kpc$^{-2}$), named supercompact UVLGs (ScUVLGs), are ideal candidates for Integral-field Spectroscopy (IFS). They are bright sources whose compact sizes (with half-light radii $\approx 1-2$ kpc) accommodate the limited fields-of-view (FOV) of many IFS instruments, typical redshifts place the Paschen-$\alpha$ (Pa-$\alpha$) line conveniently in the K-band filter, and unusually high star-formation rates (SFRs), with SFR$=$3$-$100 \msun yr$^{-1}$, designate them as a unique sample. These ScUVLGs bear striking resemblances to Lyman Break Galaxies (LBGs) -- sharing similar specific SFRs, metallicities, kinematics, and attenuations (\cite{heckman05}, \cite{choopes}, \cite{me2}).  When resimulated at high redshift, their morphologies are similar to those of LBGs \citep{rod}. ScUVLGs constitute a local (z$\sim$0.2) sample well-suited for studying details of galaxy formation with very high physical resolution and sensitivity using IFS.  

LBGs, selected by the Lyman break dropout technique \citep{steidel00}, form stars at intense rates, dominating the UV luminosity density at $z>$2. \cite{Bouwens} find only a modest decrease in the UV luminosity density out to $z=$6, indicating that LBGs represent a major phase in early stages of galaxy formation and evolution. Many studies have used IFS to observe the dynamics of LBGs, at $z=2-3$, in order to pinpoint the dominant formation process for these galaxies (i.e., \cite{Law2, Law09}, \cite{nesvadba}, \cite{FS}, \cite{wright}, \cite{nesvadba2}, and \cite{stark}). These studies probe key questions: What causes such intense star formation in these systems -- disk instabilities or tidal shocks from mergers? How does angular momentum evolve in early star-forming galaxies? Yet, at the high redshifts of LBGs, spatial resolution and depth are fundamentally limited. Except in the fortuitous but rare cases of gravitationally lensed LBGs, IFS at $z>$2 may be biased towards brighter and larger LBGs whose dynamics differ from typical LBGs \citep{nesvadba2, stark}. 

We have observed 3 ScUVLGs with the OSIRIS instrument\footnote{The data presented herein were obtained at the W.M. Keck Observatory, which is operated as a scientific partnership among the California Institute of Technology, the University of California and the National Aeronautics and Space Administration. The Observatory was made possible by the generous financial support of the W.M. Keck Foundation.}. While a larger sample of ScUVLGs has been commissioned for IFS studies, in this Letter we share initial results and compare them with high-$z$ LBG studies.  In \S\ref{sec:data}, we discuss data reduction, and analyze the results from these observations in \S\ref{sec:results}. We discuss the relevance of this work, related to other high-$z$ LBG studies in the final section, \S\ref{sec:end}. Throughout our analysis, we apply $\Lambda$CDM cosmology with (H$_0$, $\Omega_M$, $\Omega_\Lambda$)$=$(70 km s$^{-1}$Mpc$^{-1}$, 0.3, 0.7). 
\section{Data and Analysis}\label{sec:data}
Using the OSIRIS (OH Suppressing InfraRed Imaging Spectrograph; \cite{OSIRIS}) instrument at the KeckII telescope, we observed SDSS~J092600.40$+$442736.1 and SDSS~J143417.15$+$020742.5 (referred to as ScUVLG092600 and ScUVLG143417, respectively) in Spring, 2008 and SDSS~J210358.74$-$072802.4 (ScUVLG210358, henceforth) in Fall, 2008. OSIRIS utilizes the LGSAO system (Laser Guide Star Adaptive Optics; \cite{wizinowich,vandam}) to improve spatial resolution. 

\begin{figure}[t]
  \begin{center}
   \includegraphics[height=2.2in]{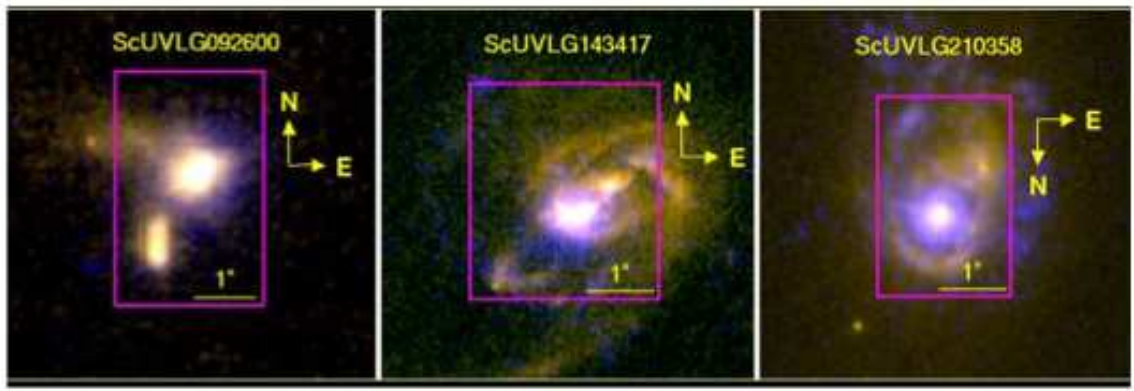}
          \caption{HST optical (red and green) $+$ UV (blue) images (left to right, see \cite{rod, rodLetter}): ScUVLG092600, ScUVLG143417, and ScUVLG210358. Magenta boxes mark OSIRIS FOV.}
   \label{fig:hst}
  \end{center}
\end{figure}

Several criteria were applied in selecting the targets and observing configuration. Using the LGSAO requires that targets be within 60\arcsec~of a bright (M$_R \lesssim$ 17 mag) tip-tilt reference star. Certain filter and spaxal scale combinations result in excess thermal background and/or detector noise \citep{Law1}. To optimize the sensitivity of our observations, we chose the Kn3 filter with 50 mas lenslet$^{-1}$ scale. For this combination, the resulting FOV for our observations was approximately 2\arcsec $\times$ 3\arcsec~(the actual FOV varied because of slight re-centering during observations -- see magenta boxes in Fig. \ref{fig:hst}). The resolution and summary of our observations for each target are given in Table \ref{tab:targets}. Our velocity resolution is approximately 34-35 km s$^{-1}$. Total on-target integration times were $\approx$ 2 hours per target.

As previously mentioned, selection of ScUVLGs depends on FUV luminosity and surface-brightness exceeding 10$^{10.3}$ L$_\odot$ and 10$^9$ L$_\odot$ kpc$^{-2}$, respectively. We exclude AGN from this sample using two separate methods: radio luminosities (see \cite{me2}; \cite{Best}), and optical emission lines (see \cite{choopes}; \cite{BPT}). The ScUVLG sample consists of compact, starburst dominated galaxies exhibiting signs of small-scale mergers or interactions only detected in the rest-frame optical at the resolution and sensitivity of HST \citep{rod}. Our 3 galaxies span stellar mass and SFR ranges of the full set of ScUVLGs. In Table \ref{tab:targets}, we list some of their physical properties. Stellar masses are determined by fitting the spectral energy distribution (SED) using unresolved Sloan Digital Sky Survey (SDSS) photometry (see \cite{Samir} and \cite{GK03b}), assuming a \cite{Kroupa} Initial Mass Function (IMF). The SFRs are unattenuated SFRs, determined from radio luminosities (see \cite{me2}). We estimate that typical uncertainties associated with the stellar mass and SFRs are $\sim$30\% based on comparing different mass modeling methods and SFR indicators. Average values and standard deviations for these properties in the full sample of 45 ScUVLGs (studied in \cite{me2}) are as follows: $\langle$SFR$\rangle = 21$ \msun yr$^{-1}$ and $\sigma_{\rm SFR} = 16$ \msun yr$^{-1}$, $\langle$log M$_*/$\msun$ \rangle = 10.0$ and $\sigma_{\rm log M_*/M_\odot} = 0.5$. While both ScUVLG143417 and ScUVLG210358 are dominated by luminous, compact UV and line emitting regions, both contain significant low surface brightness disk structure and are more massive than the typical ScUVLG. 

\begin{figure}[t]
  \begin{center}
   \includegraphics[width=6.0in]{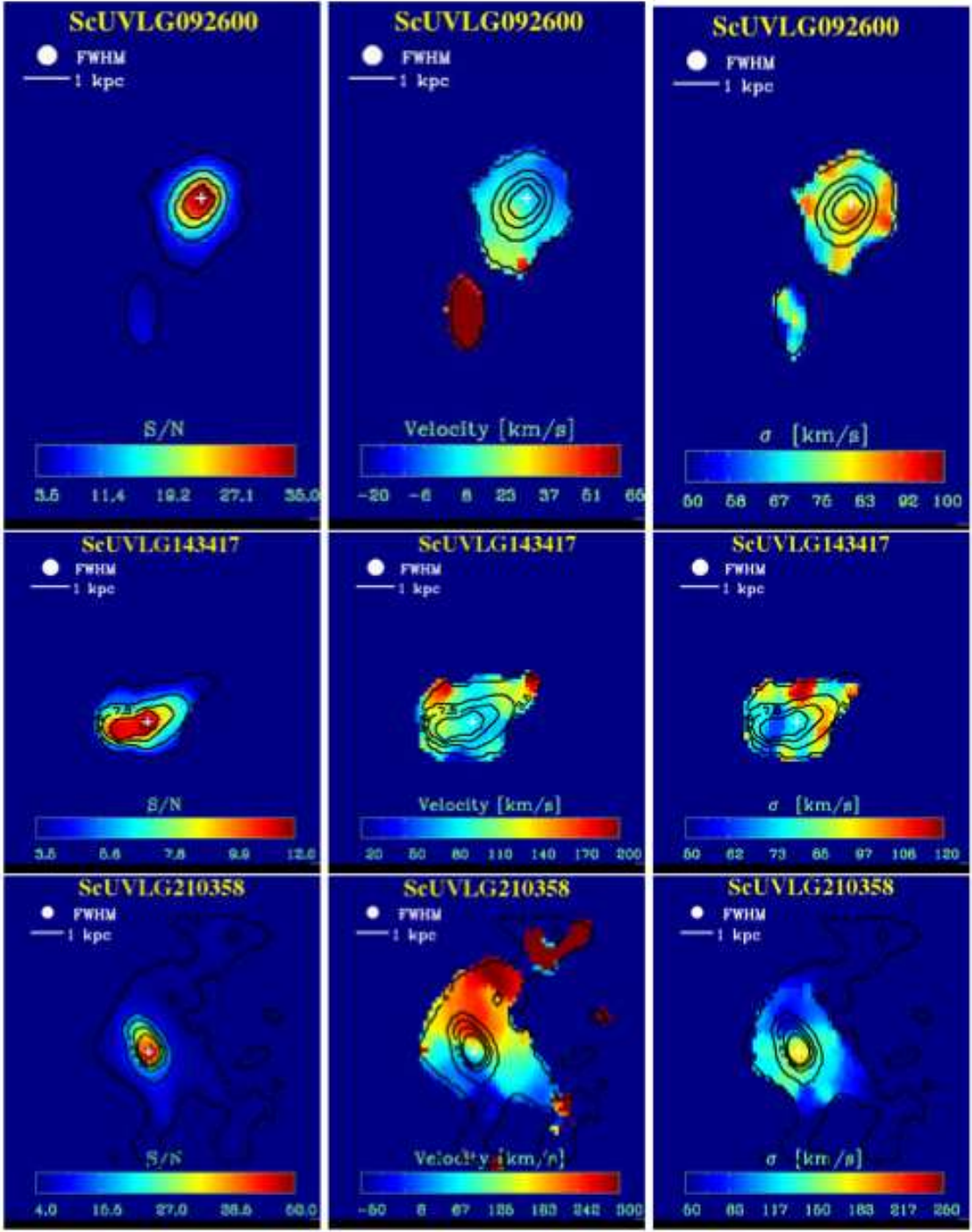} 
          \caption{OSIRIS IFS results for three ScUVLGs-- ScUVLG092600 (top), ScUVLG143417 (middle row) and ScUVLG210358(bottom). We show S/N maps (left), velocity maps (middle column), velocity dispersions (right) with S/N contours overlaid and nucleus (center of the brightest region) displayed as a white triangle. Contour levels correspond with marked values of the S/N color bar. The smoothed PSF and physical scale are shown at upper left.}
   \label{fig:maps}
  \end{center}
\end{figure}

We create 3-dimensional data cubes of spatial versus spectral dimensions by using the Keck/OSIRIS data reduction pipeline. The pipeline performs the following tasks: corrects detector non-linearity to form a data cube with calibrated wavelength units, subtracts a sky frame from the science frame to combine separate exposures into a single mosaicked data cube, removes cosmic rays, and corrects for detector cross talk. A custom IDL routine determines a mask for good data based on a signal-to-noise (S/N) threshold set at 3$\sigma$. After boxcar smoothing by a kernel of 3 lenslets in the spatial dimensions to increase the S/N, gaussians were fit to the spectral data to determine the peak, peak location, and FWHM of the Pa-$\alpha$ feature -- creating S/N maps, velocity maps and velocity dispersion ($\sigma$) maps, respectively. Resulting images for the 3 targets are shown in Fig. \ref{fig:maps} (with Pa-$\alpha$ S/N contours marked for reference).  Fig. \ref{fig:hst} displays the orientation and FOV of these targets. 

\begin{figure}[t]
  \begin{center}
   \includegraphics[width=2.0in]{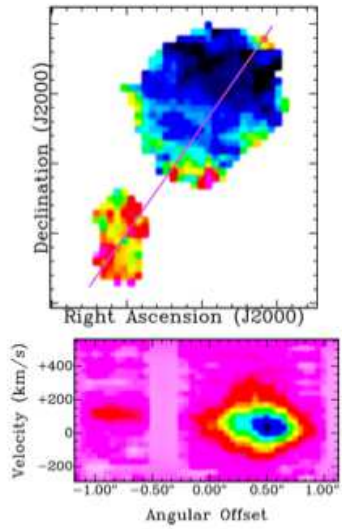} \\
       \caption{PV diagram for ScUVLG092600 is shown at bottom for a slice along the velocity gradient (marked in magenta, on the velocity field, at top). PV diagram illustrates slight velocity gradient with high dispersion.}
  \label{fig:scuvlg0926}
  \end{center}
\end{figure}

For ScUVLG092600, we also made position-velocity (PV) diagrams (shown in Fig. \ref{fig:scuvlg0926}) using the {\it kpvslice} tool in KARMA \citep{Gooch}. An integrated spectrum combining all the unmasked data was created using custom IDL software. To calculate inclination, $i$, we assume that face-on ($i=$0) disks have circular isophotes and edge-on disks have intrinsic axis ratios of 0.2 \footnote{While typical spiral galaxies have intrinsic axis ratios (q$_0$) of 0.2, ScUVLGs may be thicker. We note that even for q$_0$=0.5, values of v$_c$ and v/$\sigma$ differ by $\lesssim$ 10\%, within measurement uncertainties.}. Using Pa-$\alpha$ S/N maps, we use ratios of minor to major axes of the isophotes to measure inclinations. We use several isophote levels to determine the uncertainty of this measurement.

\section{Results}\label{sec:results}
HST images (shown in Fig. \ref{fig:hst}, see \cite{rod, rodLetter}) reveal companions and disturbed morphologies, such as tidal features and plumes. Using IFS, we analyze the velocity structure for signs of smooth rotation. In two cases, we see evidence of regular rotation (ScUVLG092600 and ScUVLG210358), while ScUVLG143417 has no clearly defined velocity structure. However, even in the cases of ordered velocity, we find high dispersion (v/$\sigma\lesssim$1), suggesting dispersion dominated systems rather than smoothly rotating thin disks. 
\subsection{ScUVLG092600}
The bright component in the NW of this system has weak, but regular velocity structure across its disk.  HST H$\alpha$ imaging reveals a filamentary component extending into the south-west, potentially suggesting an outflow. Since this feature is undetected in optical or UV morphologies, it is unlikely to be a tidal feature or spiral arm. The mild velocity gradient may arise from an outflow in the south-western region of the disk, rather than from smooth rotation. For a rotating disk scenario, we find an inclination, $i = 40 \pm 5^{\circ}$. The bottom panel of Fig. \ref{fig:scuvlg0926} displays the PV diagram, sliced along the central region of the disk (shown in magenta on the velocity field, at top). We determine a circular velocity of $v_c=\frac{20 \rm km s^{-1}}{{\rm sin} i}\approx$31$^{+4}_{-3}$ km s$^{-1}$ . Qualitatively, the PV diagram shows that velocity dispersion dominates the velocity structure across the disk. Quantitatively, the integrated spectrum gives $\sigma=$84 $\pm{3}$ km s$^{-1}$, and v$/\sigma$ for this component is 0.37$\pm0.05$. Assuming that the observed velocity dispersion of the ionized gas for this component can provide an estimate for the disk's dynamical mass, M$_{dyn}$, we use the following relation: 
\begin{eqnarray} \label{eqn:mass}
M_{\rm dyn}&=&\frac{C\sigma^2 R}{G}
\end{eqnarray}
where $C$ is the geometric constant ($C=$ 3.4 for thin rotating disks, $C=$ 5 for spherical geometry), $G$ is the gravitational constant, $R$ is the radius of the disk, and $\sigma$ is the velocity dispersion. Adopting the following values: R$=$0.3\arcsec$=$0.9 kpc, $\sigma=$84 km s$^{-1}$, we determine M$_{\rm dyn}=$4.4-6.5$\times$10$^{9}$ M$_\odot$, corresponding to 3.4$<$C$<$5. 

The companion in the south-east, spanning few velocity resolution units across, has systemic velocity of $\Delta v=+$80 km s$^{-1}$ compared to the brighter component. While we do not detect any ionized gas linking these two components, they are likely in the process of interacting given their close proximity ($\leq 3$ kpc). As further evidence of interaction, \cite{tsg} find that the bright component's central region is consistent with regular rotation, while outer edges become more disturbed and dispersion dominated.

\subsection{ScUVLG143417}
This galaxy's morphology differs from ScUVLG092600, with two bright cores (separated by $<$1 kpc) embedded in a fainter, disturbed (irregular and asymmetric) galaxy. We cannot detect the fainter arms very clearly in the Pa-$\alpha$ data. However, the central region surrounding the brightest two star-forming knots does appear kinematically disturbed -- lacking structure in velocity and velocity dispersion.

While a face-on disk could explain the absence in velocity structure, lack of structure in $\sigma$ makes that scenario less plausible. Therefore, inclination is unconstrained. Using the integrated spectrum, we find $\sigma=$86 $\pm{3}$ km s $^{-1}$. With no observed velocity gradient, we can only place an upper limit: v/$\sigma <$0.7, using v$_c=($v$_{\rm max}-$v$_{\rm min})/2$ and no inclination correction. This case represents disturbed kinematics with multiple cores and underlying faint features, suggestive of recent merger activity. 

\subsection{ScUVLG210358}
While morphologically similar to ScUVLG143417, we find that the kinematics in this galaxy are quite different. Again, the optical image reveals faint, irregular and asymmetric arms with a bright, unresolved star-forming core. However, in this case, there is a single bright, symmetric core. This galaxy is one of the most massive and luminous ScUVLGs in our sample. The position angle for the Pa-$\alpha$ isophotes (from the S/N contours) does not align with that of the velocity field, possibly indicating the presence of a warp or bar. The velocity gradient across the highest S/N region of this galaxy is very regular, maintaining similar structure to the outer, fainter regions of the system. Hints of a merger are apparent from HST images, displaying irregular, clumpy light distribution and a potential faint companion to the south-west. Yet the dynamics of this galaxy's central region show ordered rotation, with the $\sigma$ peak corresponding with the galaxy's nucleus. Therefore this galaxy has dynamics consistent with disks. We find an inclination of $i=51\pm 4^{\circ}$, circular velocity of $v_c=\frac{145 \rm km s^{-1}}{{\rm sin} i} =$186$^{+12}_{-9}$ km s$^{-1}$, and $\sigma=$175 $\pm{4}$km s$^{-1}$, giving v/$\sigma \approx$1.1. Using Eqn. \ref{eqn:mass}, at R$=$1.5kpc, we derive M$_{\rm dyn}=4 \times 10^{10}$M$_\odot$ (for C=3.4) and 6$ \times 10^{10}$M$_\odot$ (for C=5). Similar to ScUVLG092600, the signature for rotation is present, yet, this system is marginally dispersion dominated.
\subsection{Simulation to high-$z$}\label{sec:highz}
To best compare our results with high-$z$ IFS observations (tracing H$\alpha$ emission in the $K_s$-band), we perform artificial redshifting to determine how our 3 galaxies would appear if placed at $z\sim$2.  Using the recovered flux, velocity, and dispersion maps described in \S 3 we generate a synthetic model data cube which is rebinned to the angular scale which would be observed at redshift $z= 2.2645$ (chosen so that H$\alpha$ emission in $K_s$ band avoids strong OH emission features).  Assuming that H$\alpha$ emission traces Pa-$\alpha$, we normalize this model to the total H$\alpha$ luminosity determined by the unresolved SDSS spectroscopic data and artificially `observe' the model galaxy using the OSIRIS simulation code of \cite{Law1} and data reduction and analysis techniques of \cite{Law2}.  We refer the reader to these papers for a full description of the routines, noting that the simulation code fully accounts for the characteristics of the OSIRIS+LGSAO system and has been shown to well represent actual IFS observations of high-redshift galaxies \citep{Law2}.  Each of our 3 galaxies is artificially redshifted and re-observed, using a simulated 50 mas lenslet scale in $K_s$ band for a total of 2 hours of integration time under typical atmospheric conditions.  These parameters are chosen to represent typical observational values used in \cite{Law2,Law09}. 

\begin{figure}[htb]
  \begin{center}
   \includegraphics[height=7in]{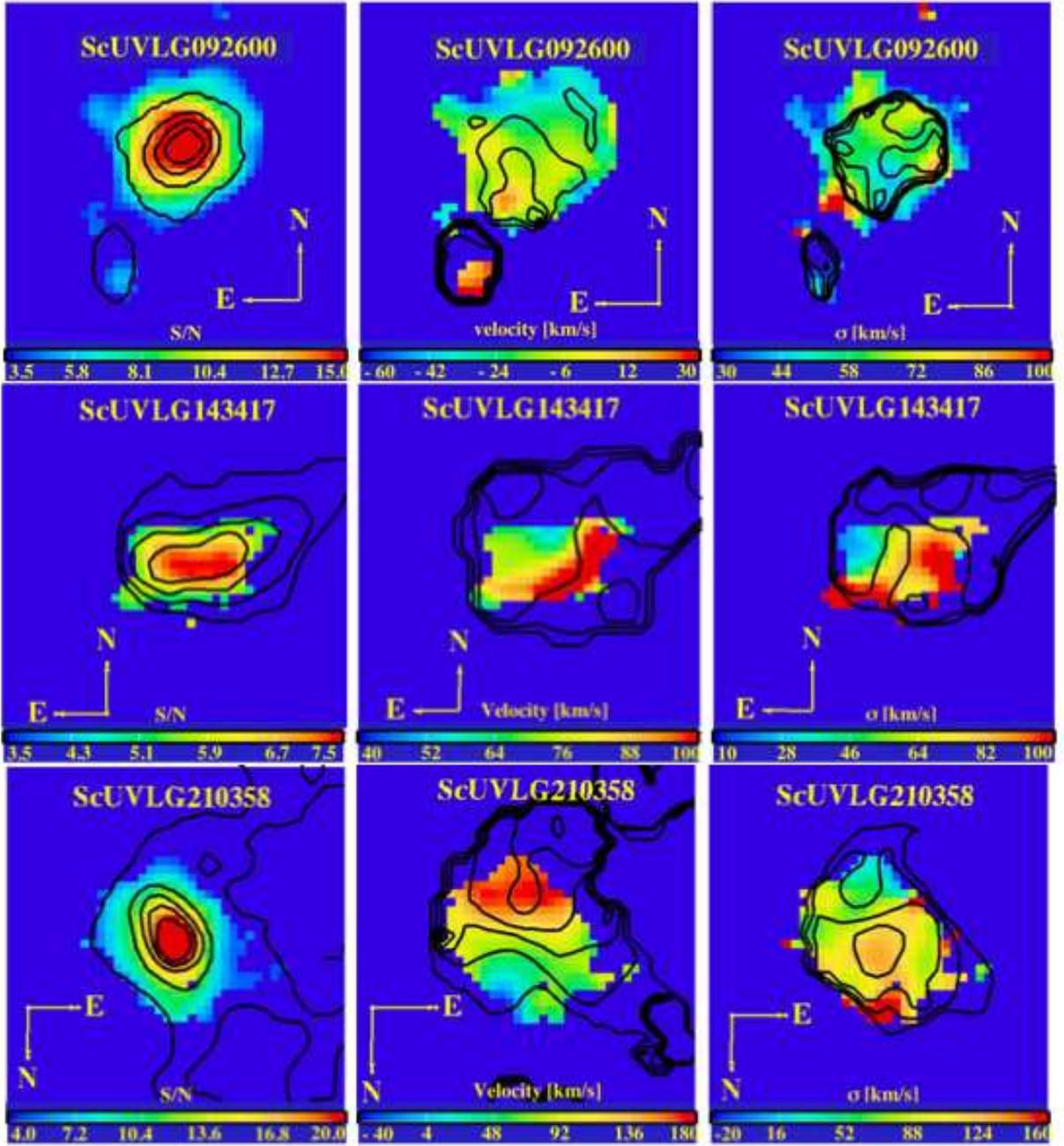} 
          \caption{High-$z$ simulations for S/N, velocity, and $\sigma$ maps (from left to right column) for ScUVLG092600 (top), ScUVLG143417 (middle row), and ScUVLG210358 (bottom). Overlying black contours outline data from Fig. \ref{fig:maps}, such that contours show low-$z$ velocity data in middle column and low-$z$ $\sigma$ data in the right column, with levels matching marked values from color bars in Fig. \ref{fig:maps}. }
   \label{fig:hiz}
  \end{center}
\end{figure}

We show the high-$z$ simulated data in Figure \ref{fig:hiz}, with black contours marking the $z\sim$0.2 data. For example, the middle panels display high-$z$ velocity with low-$z$ velocity (see data in Figure \ref{fig:maps}) contours overlaid. In all cases, the high-$z$ data matches the brightest low-$z$ features. However, this figure illustrates some of the difficulties of high-redshift LBG observations: resolution loss in the simulated S/N data causes ScUVLG092600 to appear more face-on; the simulated $\sigma$-map of ScUVLG092600 appears more disturbed as edge features dominate; lower resolution blends the double core of ScUVLG143417 into a single blob; selecting velocities where S/N exceeds detectable limits gives the velocity map in ScUVLG143417 a slightly more regular appearance; surface brightness dimming hides the complex fainter structure in ScUVLG210358. We derive values for axis ratios, inclination, circular velocity, $\sigma$ and v/$\sigma$ from the high-$z$ simulated data (shown in Table 1, alongside the observed low-$z$ results). 

The v/$\sigma$ for high-$z$ simulations of ScUVLG092600, ScUVLG143417 and ScUVLG210358 are 1.0$^{+0.3}_{-0.6}$, 0.4$\pm{0.1}$(no inclination correction), and 0.9$\pm0.1$, respectively. Therefore in all the cases, v/$\sigma$ in high-$z$ simulations differs somewhat from low-$z$ values (as shown in Table 1). For ScUVLG092600, v/$\sigma$ increases in the high-$z$ simulated case, while decreasing for the other two galaxies. We find that the high-$z$ simulations physically resemble and have similar, though slightly higher, v/$\sigma$ values as $z=$2 LBGs, which have v/$\sigma\sim$ 0.3$-$0.7 \citep{Law09}.

\section{Discussion and Summary}\label{sec:end}
We are pioneering a new study of the dynamics and kinematics in local ($z<$0.3) UV-selected galaxies that resemble LBGs in most physical properties. Similar studies have yielded mixed results about LBG dynamics at high-$z$, but our study of 3 low-$z$ ScUVLGs shows that dispersion dominated kinematics is a viable scenario. Although 2 of the 3 cases show some signs of smooth rotation, v/$\sigma\lesssim 1$ in these galaxies. Since HST images show disturbed morphologies, nearby companions, and evidence of ongoing interactions (tidal features, plumes, etc.), it is not surprising that velocity dispersions in these galaxies are so high. 

IFS studies of dynamics in LBGs draw conclusions ranging from smooth rotation to dispersion dominated kinematics. \cite{Genzel08} study a sample of massive ($M\sim10^{11}$\msun), $z=2-3$ rest frame UV-selected galaxies and classify these as turbulent, clumpy disks with formed bulges, undergoing secular evolution. Studying a subset of LBGs having the highest velocity fields, \cite{FS} find that the majority exhibit smooth rotation, consistent with cosmological predictions of flattened disk-like galaxies \citep{MMW}, while a more comprehensive sample demonstrates larger contributions from both major mergers and high dispersion systems \citep{FS09}.  \cite{nesvadba} and \cite{stark} determine gravitationally lensed LBGs at $z=$3.2 and 3.07, respectively, are examples of rotationally supported disks. \cite{wright} fit their z$=$1.5 LBG with a rotation model but cannot exclude a merger scenario. \cite{Law09} argue for largely dispersion dominated kinematics in their sample of LBGs. Analyzing IFS data for one massive LBG at $z=$3.2, \cite{nesvadba2} find evidence for dispersion dominated kinematics. 

Comparing our sample of LBG analogs to LBGs at high-$z$, our results are more consistent with compact, kinematically disturbed systems (\ie similar to \cite{Law09}, and see discussion by \cite{FS09}) rather than larger, more massive disks (\ie \cite{Genzel08}, \cite{FS}). While our selection criteria for ScUVLGs includes compactness of the UV-emitting region, HST images do reveal underlying low surface brightness, extended structure in two cases -- ScUVLG1434 and ScUVLG2103. These two examples represent the most massive ScUVLGs. Meanwhile, for high-$z$ observations, optimization of candidates may bias this study by selecting larger and brighter galaxies and may not represent the general physics governing galaxy behavior at these redshifts \citep{nesvadba2, stark}. Our view at low-$z$ may provide insight about smaller, more compact, $z=2-3$ LBGs. In these low-$z$ cases, we argue that mergers may be responsible for the observed kinematics and morphologies and speculate that a similar scenario driving LBG characteristics at high-$z$ cannot be excluded by current data (see also \cite{rod}).

\cite{Lehnert09} investigate other mechanisms capable of causing high dispersions in high-$z$ star-forming galaxies. They find that gas accretion cannot sustain the observed high dispersions. Studying the relationship between velocity dispersion and emission line characteristics, they conclude intense star formation can account for high dispersions in these systems through a feedback cycle between turbulence in the interstellar medium (ISM) and star formation. The proposed scenario may offer a plausible explanation for observed gas kinematics once galaxies reach a turbulent phase and associated starburst \citep{Lehnert09}.

We show that simulated high-$z$ data produces v/$\sigma$ values somewhat different from actual values. Misinterpretation of kinematic features may result from surface brightness and resolution effects. However, these simulations do resemble $z=$2 LBGs studied by \cite{Law09} in physical appearance and v/$\sigma$. 

Having launched this pilot study, we will continue researching a larger sample of ScUVLGs -- performing kinemetry analysis \citep{shapiro} in order to draw statistically robust conclusions about the dynamics of these low-$z$, high SFR, UV-selected galaxies (see \cite{tsg}). Detailed IFS studies of local LBG analogs offer a unique method for capturing important physics -- pinpointing the mechanisms most generally responsible for the high star-formation phase occurring in LBGs at high-$z$ and in ScUVLGs at low-$z$. \\

We thank the referee for helpful and insightful comments. The authors wish to recognize and acknowledge the very significant cultural role and reverence that the summit of Mauna Kea has always had within the indigenous Hawaiian community.  We are most fortunate to have the opportunity to conduct observations from this mountain. A.R.B. wishes to thank Dr. Jacqueline van Gorkom for her advice.

\begin{deluxetable}{llllccc|ccccc|ccccc}

\rotate
\setlength{\tabcolsep}{0.02in}
\tabletypesize{\scriptsize}
\tablecolumns{17} 
\tablewidth{0pc} 
\tablecaption{Summary of Observations} 
\tablehead{
& & & & & & & & Observed &  Data & & & & high-$z$ & Simulations & &\\
\cline{8-17}
\colhead{Target} & \colhead{RA} & \colhead{Dec} & \colhead{z} & \colhead{$\theta_{PSF}$\tablenotemark{a}} & \colhead{SFR\tablenotemark{b}} & \colhead{Log (M$_*$)} & \colhead{axis ratio} &\colhead{i} &\colhead{v$_{c}$} &\colhead{$\sigma$} &\colhead{v/$\sigma$}&\colhead{axis ratio} &\colhead{i} &\colhead{v$_{c}$} &\colhead{$\sigma$} &\colhead{v/$\sigma$}\\
& & & & \colhead{(\arcsec)} & \colhead{(M$_\odot$ yr$^{-1}$)} & \colhead{(M$_\odot$)} & & \colhead{($^\circ$)}& \colhead{(km s$^{-1}$)}& \colhead{(km s$^{-1}$)} &  & & \colhead{($^\circ$)}& \colhead{(km s$^{-1}$)}& \colhead{(km s$^{-1}$)} &
}
\startdata 
ScUVLG092600 & 141.502 & 44.460 & 0.181 & 0.101(0.208) & 8 & 9.20\tablenotemark{c} & 0.78$\pm{0.05}$& 40$\pm{5}$ & 31$^{+4}_{-3}$  & 84$\pm{3}$ & 0.37$\pm{0.05}$& 0.90$\pm{0.05}$ & 26$\pm{4}$ & 59$^{+11}_{-7}$ & 59$\pm{17}$ & 1.0 $^{+0.3}_{-0.6}$\\
ScUVLG143417 & 218.571 & 2.128 & 0.180 & 0.098(0.197) & 15 & 10.89\tablenotemark{c}  & \nodata & \nodata & 60 & 86 $\pm{3}$ & $<$0.70 & \nodata & \nodata & 35 & 98$\pm{12}$ & $<$0.35 \\
ScUVLG210358  & 315.995 & -7.467 &  0.137& 0.065(0.144) & 68 & 11.05\tablenotemark{d} & 0.65$\pm{0.05}$& 51$\pm{4}$ & 186$^{+12}_{-9}$ & 175$\pm{4}$ & 1.1$\pm{0.1}$ &0.79$\pm{0.02}$& 39$\pm{2}$& 167$^{+8}_{-7}$&175$\pm{11}$& 0.9$\pm{0.1}$
 \enddata 
 \tablenotetext{a}{$\theta_{PSF}$ gives observed FWHM of the PSF, spatially smoothed FWHM is given in parentheses.}
 \tablenotetext{b}{SFR was determined from 1.4GHz-derived SFR -- proxy for total, unattenuated SFR (see \cite{me2})}
 \tablenotetext{c}{M$_*$ based on SED-fitting (see \cite{Samir}). }
 \tablenotetext{d}{M$_*$ is dust corrected median stellar mass \citep{GK03b} from MPA/JHU value added catalog (see http://www.mpa-garching.mpg.de/SDSS/index\_old.html). }
 \label{tab:targets}
\end{deluxetable} 

\end{document}